\documentclass[aps,prb,preprint,groupedaddress]{revtex4}

\usepackage[dvips]{graphicx}

\usepackage{enumerate}

\usepackage{amsmath,amssymb}

\newcommand{ \BE }{ \begin{equation} }
\newcommand{ \EE }{ \end{equation} }

\newcommand{ \BEA }{ \begin{eqnarray} }
\newcommand{ \EEA }{ \end{eqnarray} }

\newcommand{ \BSUB }{ \begin{subequations} }
\newcommand{ \ESUB }{ \end{subequations} }

\newcommand{ \Eq  }[1]{Eq.~(\ref{Eq:#1})}
\newcommand{ \Eqs }[2]{Eqs.~(\ref{Eq:#1}) and (\ref{Eq:#2})}

\newcommand{ \Sec }[1]{Sec.~\ref{Sec:#1}}
\newcommand{ \Secs }[2]{Secs.~\ref{Sec:#1} and \ref{Sec:#2}}

\newcommand{ \Ref  }[1]{Ref.~\onlinecite{#1}}
\newcommand{ \Refs }[2]{Refs.~\onlinecite{#1} and \onlinecite{#2}}

\newcommand{ \etal }{\textit{et al.}}

\newcommand{ \Cff }{ C_\mathrm{ff} }
\newcommand{ \Cffddot }{ \ddot{C}_\mathrm{ff} }

\newcommand{ \Cdd }{ C_\mathrm{dd} }
\newcommand{ \Cddddot }{ \ddot{C}_\mathrm{dd} }

\newcommand{ \Cbarff }{ \bar{C}_\mathrm{ff} }

\newcommand{ \Cffbar }{ \bar{C}_\mathrm{ff} }

\newcommand{ \tr }{ \mathrm{tr} }

\newcommand{ \Hop }{ \hat{H} }
\newcommand{ \Fop }{ \hat{F} }
\newcommand{ \xop }{ \hat{x} }
\newcommand{ \pop }{ \hat{p} }
\newcommand{ \Top }{ \hat{T} }
\newcommand{ \Vop }{ \hat{V} }

\newcommand{ \xds }{ x^\ddagger }
\newcommand{ \xref }{ x^{*} }

\newcommand{ \Rkin }{ R_\mathrm{kin} }
\newcommand{ \Rpot }{ R_\mathrm{pot} }

\begin{document}




\title{Path-integral virial estimator for reaction rate
calculation \\based on the quantum instanton approximation}







\author{Sandy Yang, Takeshi Yamamoto, and William H. Miller}

\email{miller@cchem.berkeley.edu}

\affiliation{Department of Chemistry and Kenneth S. Pitzer Center for Theoretical Chemistry, 
University of California, and Chemical Sciences Division, Lawrence Berkeley National Laboratory, 
Berkeley, California 94720}





\begin{abstract}
The quantum instanton approximation is a type of quantum transition state
theory that calculates the chemical reaction rate using the reactive flux 
correlation function and its low order derivatives at time zero. Here we present
several path-integral estimators for the latter quantities, which
characterize the initial decay profile of the flux correlation function.
As with the internal energy or heat capacity calculation,
different estimators yield different variances (and therefore different convergence
properties) in a Monte Carlo calculation. Here we obtain a virial(-type) estimator
by using a coordinate scaling procedure rather than integration by parts,
which allows more computational benefits.
We also consider two different methods for treating the flux operator,
i.e., local-path and global-path approaches, in which the latter achieves
a smaller variance at the cost of using second-order potential derivatives.
Numerical tests are performed for a one-dimensional
Eckart barrier and a model proton transfer reaction in a polar solvent,
which illustrates the reduced variance of the virial estimator
over the corresponding thermodynamic estimator.
\end{abstract}


\pacs{}

\maketitle




\section{Introduction}

Developing an accurate and practical method for computing chemical reaction
rates is one of the fundamental subjects of theoretical chemistry. In this regard
the most successful approach is probably classical transition state theory
(TST),\cite{Eyring,Wigner,TSTreview}
which has been applied widely to numerous reactions including
biological systems such as enzyme catalysis.\cite{TSTenzyme}
The robustness of TST comes from its
simplicity, i.e., the rate is determined solely from the free energy
difference between the reactant and the activated complex. TST relies on
the assumption of no ``recrossing'' trajectories through the dividing
surface, which is usually valid at not too high temperature and for
large dimensional systems. While successful in many cases, TST has the
inherent deficiency that it accounts for no quantum effects,
which needs to be addressed in order to treat
low-temperature or light-atom transfer reactions. A conventional remedy to
this problem is to add quantum corrections in a posteriori manner, e.g., by
multiplying a tunneling factor that is computed along a prescribed tunneling
path.\cite{TunnelingCorrection}
Another strategy is to try to develop a quantum TST (QTST) by starting
from the rigorous quantum rate expression and make some approximations for
neglecting recrossing effects.
Several such theories exist,\cite{Voth,Jang,Miller_QTST,Pollak,Pollak2,Yamashita,Tromp,Hansen,Krilov,Sim} 
though there is in principle no unique formulation (in contrast to the classical case).

The quantum instanton (QI) approximation \cite{QI_org,QI_scheme,QI_CH4,QI_Michele,
QI_Charu,QI_extension,QI_borgis,QI_KIE,QI_pentadiene}
is a recently developed theory for chemical reaction rates that is among the category of QTST. 
While the original derivation was based on the semiclassical ``instanton''
(periodic orbit in imaginary time) model,\cite{QI_org}
the working rate expression can be understood roughly as the
second-order cumulant (or Gaussian) approximation to the flux-flux correlation
function,
\BE
\label{Eq:cumulant}
  \Cff( t ) = \Cff( 0 ) + \frac{1}{2} \Cffddot( 0 ) t^2 + \cdots
  \simeq
  \Cff( 0 )
  \exp \left[
          \frac{1}{2} \frac{ \Cffddot( 0 ) }{ \Cff( 0 ) } t^2
       \right]
  ,
\EE
for which the rate constant is given by
\BE
\label{Eq:QIrateCff}
  k( T ) = \frac{1}{Q_r} \int_0^{\infty} dt \Cff ( t )
  \simeq
  \frac{ \Cff(0) }{ Q_r }
  \sqrt{ \frac{\pi}{2} }
  \frac{ 1 }{ \sqrt{ -\Cffddot( 0 ) / \Cff( 0 ) } }
\EE
(see \Sec{QI} for details).
This approximation can be viewed as a quantum analog of the (classical) TST
assumption in the sense that all possible
oscillations in $\Cff(t)$ at later times (quantum re-crossing flux) are neglected.
Test calculations
show that this QI approximation gives a rate accurate to within $\sim$10 \%
of the exact rate when the reaction is ``direct'', and also to
within a factor of 2 even for cases in which significant recrossing is expected
(e.g., the collinear Cl+HCl reaction).\cite{QI_Michele}
The computational merit of \Eq{QIrateCff}
is that it is expressed wholly in terms of the Boltzmann
operator, and thus it can be evaluated rigorously even for complex molecular
systems using imaginary-time path integrals. A previous paper has presented
such a scheme,\cite{QI_scheme}
in which the factor $\Cff(0)/Q_r$ in \Eq{QIrateCff}
is evaluated as the
barrier height of a particular free energy surface, while the remaining factor
is calculated as the statistical average of some estimating function over the
transition-state path ensemble. This computational scheme has been applied
successfully to several benchmark systems including gas-phase reactions such
as $\rm CH_4 + H \rightarrow CH_3 + H_2$,\cite{QI_CH4}
a model proton transfer reaction in a polar solvent,\cite{QI_borgis}
and an isomerization reaction of pentadiene.\cite{QI_pentadiene}

The purpose of this paper is to present an improved path-integral estimator
for computing the QI rate. In particular, we focus on the factor
$\Cffddot(0)/\Cff(0)$ in \Eq{QIrateCff}
which characterizes the initial decay profile of
$\Cff( t )$. This quantity involves several different estimators because of
the presence of the second time derivative. The
estimator used in previous work was of ``thermodynamic'' type,\cite{QI_scheme}
and its variance thus grows rapidly
as a function of the number of path variables employed in the path integration. As with
the internal energy or heat-capacity 
calculation,\cite{Barker,Herman,Parrinello,Cao,Janke,Glaesemann,Neirotti,PredescuVirial,YamamotoVirial}
it should be possible to transform the thermodynamic estimator
into a virial form in order to reduce the statistical error. In this paper we
present such a scheme based on a coordinate scaling procedure, rather than
integration by parts,
which is based on the recent study
by Predescu \etal{}\cite{PredescuVirial,PredescuRate}
and possesses the following computational benefits:
(i) the transformation to a virial estimator
is quite straightforward in contrast to integration by parts;
(ii) one can use a finite-difference technique in order to
avoid explicit calculation of potential derivatives in the virial
estimator; and (iii) higher-order time derivatives of $\Cff(t)$ such as
$\Cff^{(4)}(0)$, $\Cff^{(6)}(0)$, ..., can also be generated
with little modifications to the code,
which can be used as input for more flexible
approximations\cite{QI_extension} to the true $\Cff(t)$ than \Eq{cumulant}.

The remainder of this paper is as follows: In \Sec{QI} we summarize the working
expression of the QI theory. In \Sec{offdiag} we first consider an ``off-diagonal''
average energy and derive its thermodynamic and virial estimators to describe
the basic idea of coordinate scaling. In \Secs{corrfunc}{flux} we apply the
scaling procedure to quantum time correlation functions in order to obtain a virial
estimator for the reaction rate. In \Sec{test} we calculate the variance
of the virial estimator for a one-dimensional Eckart barrier and a
model proton transfer reaction in a polar solvent. \Sec{concl} concludes.


\section{The quantum instanton approximation for chemical reaction rates
\label{Sec:QI}}

The QI theory approximates the reaction rate as follows (see \Ref{QI_org}
for the derivation motivated by semiclassical considerations):
\BE
\label{Eq:QIrateOriginal}
  k( T ) \simeq \frac{ \Cff(0) }{ Q_r }
  \frac{ \sqrt{\pi} }{ 2 }
  \frac{\hbar}{\Delta H}
  ,
\EE
where $\Cff(0)$ is the zero time value of the flux-flux correlation
function,\cite{MST}
\BE
  \Cff( t ) = \tr
  \left[
     e^{- \beta \Hop / 2} \Fop e^{- \beta \Hop / 2}
     e^{i \Hop t / \hbar} \Fop e^{- i \Hop t / \hbar}
  \right]
\EE
with $\Fop$ being the flux operator,
\BE
\label{Eq:flux_op}
  \Fop
  = \frac{i}{\hbar} [ \Hop, h( \xop - \xds ) ]
  = \frac{1}{2 m} [ \pop \delta( \xop - \xds ) + \delta( \xop - \xds ) \pop ]
  ,
\EE
and $\Delta H$ represents a specific type of energy variance (i.e., $\Delta
H^2 = \langle \Hop^2 \rangle - \langle \Hop \rangle^2$),
\BE
\label{Eq:DeltaH}
  \Delta H^2
  =
  \frac
  {
    \langle \xds | \Hop^2 e^{- \beta \Hop / 2} | \xds \rangle
  }{
    \langle \xds | e^{- \beta \Hop / 2} | \xds \rangle
  }
  -
  \left[
  \frac{
    \langle \xds | \Hop e^{- \beta \Hop / 2} | \xds \rangle
  }{
    \langle \xds | e^{- \beta \Hop / 2} | \xds \rangle
  }
  \right]^2
  .
\EE
In this paper we consider a one-dimensional system with the Hamiltonian $H =
p^2 / 2 m + V ( x )$ for notational simplicity. In \Eq{flux_op}, $\xds$ is the
location of the dividing surface that separates the reactant and product
regions. $\Delta H$ in \Eq{DeltaH} can be written more compactly as
\BE
\label{Eq:DeltaH_Cdd}
  \Delta H^2
  = - \frac{\hbar^2}{2} \frac{ \Cddddot(0) }{ \Cdd(0) },
\EE
where $\Cdd(t)$ is a ``delta-delta'' correlation function defined by
\BE
\label{Eq:Cdd}
  \Cdd( t )
  =
  \tr
  \left[
  e^{- \beta \Hop / 2} \delta( \xop - \xds )
  e^{- \beta \Hop / 2} e^{i \Hop t / \hbar} \delta( \xop - \xds )
  e^{- i \Hop t / \hbar}
  \right]
  .
\EE
Substituting \Eq{DeltaH_Cdd} into \Eq{QIrateOriginal} gives
\BE
\label{Eq:QIrateCdd}
  k( T )
  \simeq \frac{ \Cff(0) }{ Q_r }
  \sqrt{ \frac{\pi}{2} }
  \frac{1}{ \sqrt{ -\Cddddot(0) / \Cdd(0) } }
  ,
\EE
which has a formal resemblance to the Gaussian approximation to
$\Cff(t)$ in \Eq{cumulant}. An extended version of the QI theory has also
been proposed,
which makes a log-augmented cumulant expansion of
$\Cff(t)$ as follows,\cite{QI_extension}
\BE
\label{Eq:EQI}
  \Cff( t )
  \simeq
  \Cff( 0 ) \exp
  \left\{
    b_0 \ln
    \left[
      1 + \left( \frac{2 t}{\hbar \beta} \right)^2
    \right]
    + b_1 t^2 + \cdots b_N t^{2 N}
  \right\}
  ,
\EE
where coefficients $\{b_k\}$ are determined by a matching procedure with the
direct Taylor series expansion of $\Cff( t )$. We note that the above approximation
still falls among QTST because $\Cff(t)$ in \Eq{EQI} is always positive and thus
does not describe any recrossing effects (see \Ref{QI_extension}
for how this extension improves upon the QI rate).


\section{Path-integral estimators for off-diagonal average energy
\label{Sec:offdiag}}

Before proceeding, it is useful first to
present the coordinate scaling idea in its simplest form by considering
an ``off-diagonal'' average energy defined by
\BE
  E_{ba} ( \beta )
  = \frac{ \langle x_b | \Hop e^{- \beta \Hop} | x_a \rangle }
         { \langle x_b | e^{- \beta \Hop} |x_a \rangle }
  =
  - \frac{\partial}{\partial \beta} \ln \rho_{ba} ( \beta )
  ,
\EE
where $\rho_{ba} ( \beta ) = \langle x_b |e^{- \beta \Hop} |x_a \rangle$,
since this quantity serves as the basis for treating a time
correlation function. Using the primitive approximation to the
Boltzmann operator,
\BE
  e^{- \beta \Hop}
 \simeq \left( e^{- \epsilon \Vop / 2} e^{- \epsilon \Top}
 e^{- \epsilon \Vop / 2} \right)^N
\EE
with $\epsilon = \beta / N$,
a discretized path integral for $\rho_{ba}$ is obtained as
\BE
\label{Eq:rho_ba}
  \rho_{ba} ( \beta ) = \int d x_1 \cdots \int d x_{N - 1}
   W_{ba} ( x_1, \ldots, x_{N - 1} ; \beta )
  ,
\EE
where
\BE
  W_{ba} ( x_1, \ldots, x_{N-1} ; \beta )
  =
  \left( \frac{m N}{2 \pi \hbar^2 \beta} \right)^{N / 2}
  \exp \left[
         - \frac{m N}{2 \hbar^2 \beta}
         \sum^N_{k = 1} ( x_k - x_{k - 1} )^2
         - \frac{\beta}{N} \sum_{k = 0}^N w_k V( x_k )
       \right]
\EE
with $x_0 = x_a$, $x_N = x_b$, and $w_k = 1 / 2$
for $k = 0, N$ and $w_k = 1$
otherwise. Differentiating \Eq{rho_ba} with respect
to $\beta$ gives a thermodynamic estimator for the energy,
\BE
  E_{ba} ( \beta ) = \left\langle \epsilon_T \right\rangle_{ba}
\EE
with
\BE
\label{Eq:thermo_ba}
  \epsilon_T = \frac{N}{2 \beta}
  - \frac{m N}{2 \hbar^2 \beta^2}
  \sum^N_{k = 1} ( x_k - x_{k - 1} )^2
  + \frac{1}{N} \sum_{k = 0}^N w_k V ( x_k )
  ,
\EE
where $\langle \cdots \rangle_{ba}$ denotes an ensemble average over the
weight function $W_{ba} ( x_1, \ldots, x_{N - 1} ; \beta )$. This
estimator has the well-known drawback that the statistical error grows with
$N$ due to cancellation of the first two terms in the right-hand side of
\Eq{thermo_ba}. As in the case of the internal energy or
heat capacity, one can transform the above estimator into a virial form through
integration by parts. Here instead we employ a coordinate scaling
procedure that we find more useful.\cite{PredescuVirial,PredescuRate,YamamotoVirial}
To this end we first
write the density matrix at a different temperature $\beta'$,
\BE
\label{Eq:rho_ba_prime}
  \rho_{ba} ( \beta' ) = \int d x_1' \cdots \int d x_{N - 1}'
  W_{ba} ( x_1', \ldots, x_{N - 1}' ; \beta' )
  ,
\EE
and then transform the integration variables $\{x_k'\}$ into a set of new variables
$\{x_k\}$ according to
\BE
\label{Eq:coord_trans}
  x_k' = \xref_k + \sqrt{\frac{\beta'}{\beta}} ( x_k - \xref_k )
  ,
\EE
where $\xref_k$ is the reference point given by
\BE
  \xref_k = x_a + ( x_b - x_a ) \frac{k}{N}
  .
\EE
Using the following identity (or with the method described in Appendix),
\BE
  \frac{1}{\beta'}
  \sum_{k = 1}^N ( x_k' - x_{k - 1}' )^2 = \frac{1}{\beta}
  \sum_{k = 1}^N ( x_k - x_{k - 1} )^2 +
  \left(
    \frac{1}{\beta'} - \frac{1}{\beta}
  \right)
  \frac{( x_b - x_a )^2}{N}
  ,
\EE
one can rewrite \Eq{rho_ba_prime} as
\BE
\label{Eq:rho_ba_scaled}
  \rho_{ba} ( \beta' ) = \int d x_1 \cdots \int d x_{N - 1}
  W_{ba} ( x_1, \ldots, x_{N - 1} ; \beta ) R_{ba} ( \beta' )
  ,
\EE
where
\BE
\label{Eq:R_ba}
  R_{ba} ( \beta' )
  = \frac{ \langle x_b | e^{- \beta' \Top} | x_a \rangle }
         { \langle x_b | e^{- \beta  \Top} | x_a \rangle }
  \exp
  \left\{ - \frac{1}{N} \sum_{k = 0}^N w_k
    \left[ V ( x_k' ) - V ( x_k ) \right]
  \right\}
  .
\EE
Note that all the $\beta'$ dependence is now embedded in the
$R_{ba}$ factor. Differentiating \Eq{rho_ba_scaled}
with respect to $\beta'$ and taking the limit $\beta' \rightarrow \beta$ gives
a virial estimator
\BE
  E_{ba} ( \beta )
  =
  \left\langle \epsilon_V \right\rangle_{ba}
  =
  \left\langle
    \left.
      \frac{\partial R_{ba} ( \beta' )}{\partial \beta'}
    \right|_{\beta' = \beta}
  \right\rangle_{ba}
\EE
with
\BE
\label{Eq:virial_ba}
  \epsilon_V
  =
  \frac{1}{2 \beta} - \frac{m}{2 \hbar^2 \beta^2} ( x_b - x_a )^2
  + \frac{1}{N} \sum_{k = 0}^N w_k
  \left[
    \frac{1}{2} ( x_k - \xref_k) V' ( x_k ) + V ( x_k )
  \right]
  .
\EE
Alternatively, one may evaluate the virial estimator via finite difference
as\cite{PredescuVirial}
\BE
  \epsilon_V
  \simeq
  \frac{ R_{ba} ( \beta + \delta \beta ) - R_{ba} ( \beta - \delta \beta )}
       { 2 \delta \beta }
  ,
\EE
in order to avoid explicit calculation of the potential derivatives.


\section{Virial estimator for the time derivative of correlation functions
\label{Sec:corrfunc}}

With the scaling procedure above it is now straightforward to derive a virial
estimator for the time derivative of correlation functions such as
$\Cffddot ( 0 )$ and $\Cddddot ( 0 )$. We start with
the following correlation function,
\BE
  C( t ) = \tr
  \left[
    e^{- \beta \Hop / 2} A( \xop ) e^{- \beta \Hop / 2}
    e^{i \Hop t / \hbar} B( \xop ) e^{- i \Hop t / \hbar}
  \right]
  ,
\EE
where $\hat{A}$ and $\hat{B}$ are
arbitrary position-dependent operators [note that $C (t)$
becomes the delta-delta correlation function in \Eq{Cdd} if we set $A ( x ) =
B ( x ) = \delta ( x - \xds )$]. For simplicity we work with the
imaginary-time counterpart,
\BE
  \bar{C} ( \lambda ) \equiv C ( - i \hbar \lambda )
  = \tr
  \left[
    e^{- ( \beta / 2 + \lambda ) \Hop} A( \xop )
    e^{- ( \beta / 2 - \lambda ) \Hop} B( \xop )
  \right]
  ,
\EE
with which the time derivative is given by $( d / d t )^n C ( 0 )
= ( i / \hbar )^n ( d / d \lambda )^n \bar{C} ( 0 )$. Discretizing the
Boltzmann operators $\exp [ - ( \beta / 2 \pm \lambda ) \Hop ]$ with $P / 2$ time
slices gives
\BE
\label{Eq:C_original}
  \bar{C} ( \lambda ) = \int d x_1 \cdots \int d x_P A ( x_0 ) B ( x_{P / 2} )
  W ( x_1, \ldots, x_P ; \lambda )
  ,
\EE
where
\BE
  W ( x_1, \ldots, x_P ; \lambda )
  =
  \left[
    \frac{m P}{2 \pi \hbar^2 ( \beta + 2 \lambda )}
  \right]^{P / 4}
  \left[
    \frac{m P}{2 \pi \hbar^2 ( \beta - 2 \lambda )}
  \right]^{P / 4}
  \exp(- S)
\EE
and
\BEA
  S
  & = &
  \frac{m P}{2 \hbar^2 ( \beta + 2 \lambda )}
  \sum^{P / 2}_{k = 1} ( x_k - x_{k - 1} )^2
  + \frac{1}{P} \sum^{P / 2}_{k = 0} \tilde{w}_k ( \beta + 2 \lambda ) V ( x_k )
  \nonumber
  \\
  &  &
  + \frac{m P}{2 \hbar^2 ( \beta - 2 \lambda )}
  \sum^P_{k = P / 2 + 1} ( x_k - x_{k - 1} )^2
  + \frac{1}{P} \sum^P_{k = P / 2} \tilde{w}_k ( \beta - 2 \lambda ) V ( x_k )
\EEA
with $x_0 = x_P$ and $\tilde{w}_k = 1 / 2$ for $k = 0, P / 2, P$
and $\tilde{w}_k = 1$ otherwise.
Differentiating $\bar{C} ( \lambda )$ in \Eq{C_original} with respect to
$\lambda$ and taking the $\lambda \rightarrow 0$ limit gives a thermodynamic
estimator for $\ddot{C} ( 0 )$ (note that the first
derivative vanishes by symmetry):
\BE
  \frac{\ddot{C} ( 0 )}{C( 0 )}
  = - \frac{1}{\hbar^2}
  \left\langle
    F_T^2 + G_T
  \right\rangle
  ,
\EE
where
\BE
  F_T = \frac{m P}{\hbar^2 \beta^2}
  \left(
    \sum^{P / 2}_{k = 1} - \sum^P_{k = P / 2 + 1}
  \right)
  ( x_k - x_{k - 1} )^2 - \frac{2}{P}
  \left(
    \sum_{k = 1}^{P / 2 - 1} - \sum_{k = P / 2 + 1}^{P - 1}
  \right)
  V ( x_k )
\EE
and
\BE
  G_T = \frac{2 P}{\beta^2}
  - \frac{4 m P}{\hbar^2 \beta^3} \sum_{k = 1}^P ( x_k -  x_{k - 1} )^2
\EE
with $\langle \cdots \rangle$ denoting an ensemble average over the weight
function $A ( x_0 ) B ( x_{P / 2} ) W ( x_1, \ldots, x_P ; 0 )$. 
This is the estimator that has been employed in previous
work.\cite{QI_scheme,QI_CH4,QI_borgis,QI_KIE}
To transform it into virial form,
we write $\bar{C} ( \lambda )$ in terms of temporary variables $\{x_k'\}$,
\BE
  \bar{C}( \lambda ) = \int d x_1' \cdots \int d x_P' A ( x_0' ) B ( x_{P / 2}' )
  W ( x_1', \ldots, x_P' ; \lambda )
  ,
\EE
and introduce a set of new variables $\{x_k\}$ as follows:
\BE
  x_k' =
  \begin{cases}
    \xref_k + \sqrt{ \frac{\beta + 2 \lambda}{\beta} } ( x_k - \xref_k )
    &
    \quad ( 0 < k < P / 2 )
    \\
    \xref_k + \sqrt{\frac{\beta - 2 \lambda}{\beta}} ( x_k - \xref_k ) 
    &
    \quad ( P / 2 < k < P )
    \\
    x_k
    &
    \quad ( k = 0, P/2, P )
  \end{cases}
\EE
with
\BE
  \xref_k = \xref_{P - k} = x_0 + ( x_{P / 2} - x_0 ) \frac{k}{P / 2}
  \quad ( 0 \leq k \leq P / 2 )
  .
\EE
The expression for $\bar{C}( \lambda )$ then becomes
\BE
  \bar{C} ( \lambda ) = \int d x_1 \cdots \int d x_P A ( x_0 ) B ( x_{P / 2} )
  W ( x_1, \ldots, x_P ; 0 ) R ( \lambda )
  ,
\EE
where $R( \lambda ) = \Rkin \Rpot$ with
\BE
  \Rkin
  =
  \frac{ \langle x_P | e^{- ( \beta / 2 - \lambda ) \Top} | x_{P / 2} \rangle
         \langle x_{P / 2} | e^{- ( \beta / 2 + \lambda ) \Top} | x_0 \rangle
       }
       { \langle x_P | e^{- \beta \Top / 2} | x_{P / 2} \rangle
         \langle x_{P / 2} | e^{- \beta \Top / 2} | x_0 \rangle}
\EE
and
\BEA
  \Rpot
  & = &
  \exp
  \left\{
    - \frac{1}{P} \sum_{k = 0}^{P / 2} \tilde{w}_k
    \left[
      ( \beta + 2 \lambda ) V ( x_k' ) - \beta V ( x_k )
    \right]
  \right.
  \\
  &  &
  \left.
    - \frac{1}{P} \sum_{k = P / 2}^P \tilde{w}_k
    \left[
      ( \beta - 2 \lambda ) V ( x_k' ) - \beta V ( x_k )
    \right]
  \right\}
  .
\EEA
Differentiating this expression for $\bar{C}(\lambda)$ with respect to
$\lambda$ gives the desired virial estimator,
\BE
  \frac{\ddot{C} ( 0 )}{C_{} ( 0 )} = - \frac{1}{\hbar^2}
  \left\langle
    F_V^2 + G_V
  \right\rangle
\EE
with
\BE
  F_V = - \frac{2}{P}
  \left(
    \sum_{k = 1}^{P / 2 - 1}
    - \sum_{k = P / 2 + 1}^{P - 1}
  \right)
  \left[
    \frac{1}{2} ( x_k - \xref_k ) V' ( x_k ) + V ( x_k )
  \right]
\EE
and
\BE
  G_V = \frac{4}{\beta^2} - \frac{16 m}{\hbar^2 \beta^3}
  ( x_0 - x_{P / 2} )^2 - \frac{1}{\beta P}
  \sum_{k = 1}^P
  \left[
    3 ( x_k - \xref_k ) V' ( x_k ) + ( x_k - \xref_k )^2 V'' ( x_k )
  \right]
  .
\EE
In practice we can avoid the calculation of first- and second-order potential
derivatives by numerically differentiating $R(\lambda)$ as
\BE
  \frac{\ddot{C} ( 0 )}{C( 0 )}
  \simeq - \frac{1}{\hbar^2}
  \left\langle
    \frac{R ( \delta \lambda ) + R ( - \delta \lambda ) - 2 R ( 0 )}
         { ( \delta \lambda )^2 }
  \right\rangle
  .
\EE


\section{Treatment of the flux operator
\label{Sec:flux}}

\subsection{Local-path approach
\label{Sec:path_local}}

Applying the above scheme to the flux-flux correlation function is
somewhat tricky because of the nonlocal character of the flux operator (i.e.,
a derivative operator). Different estimators arise depending on the route of
the derivation, which in general exhibit different magnitudes of the variance.
In previous work\cite{QI_scheme,QI_CH4,QI_borgis,QI_KIE}
we have employed a ``local-path'' estimator, in which the
flux operator was evaluated in terms of a few path variables near the dividing
surface. This local estimator can be combined with the coordinate scaling
procedure as follows. First we construct a discretized path integral for
$\Cbarff ( \lambda ) = \Cff ( - i \hbar \lambda )$ as in
\Sec{corrfunc}, in which the following matrix element appears:
\BE
   K_\mathrm{fi}
   =
   \langle
     x_1' | e^{- ( \beta + 2 \lambda ) \Hop / P} \Fop
            e^{- ( \beta - 2 \lambda ) \Hop / P} | x_{- 1}'
   \rangle
   ,
\EE
where $\{x_k'\}$ are temporary variables to be scaled later. Making the primitive
approximation to $e^{- ( \beta \pm 2 \lambda ) \Hop / P}$
and evaluating the flux operator analytically via \Eq{flux_op} gives
\BE
  K_{\mathrm{fi}}
  \simeq
  \int d x_0' \delta ( x_0' - \xds ) v_0 ( \lambda )
  \langle
    x_1' | e^{- ( \beta + 2 \lambda ) \Hop / P} | x_0'
  \rangle
  \langle
    x_0' | e^{- ( \beta - 2 \lambda ) \Hop / P} | x_{-1}'
  \rangle
  ,
\EE
where the velocity factor $v_0 ( \lambda )$ is defined by
\BE
\label{Eq:velfac}
  v_k ( \lambda )
  =
  \frac{i P}{2 \hbar}
  \left(
    \frac{x_{k + 1}' - x_k'}{\beta + 2 \lambda}
     + \frac{x_k' - x_{k - 1}'}{\beta - 2 \lambda}
  \right)
   -
  \frac{i \hbar \lambda}{m P} V' ( x_k' )
  \EE
with $k = 0$. The effect of the flux operator is thus expressed in terms of
only three path variables. Treating another flux operator in
$\bar{C}_{\mathrm{ff}} ( \lambda )$ with the same method and performing the
coordinate scaling precisely as in the preceding section gives
\BE
  \Cffbar ( \lambda ) = \int d x_1 \cdots \int d x_P
  W ( x_1, \ldots, x_P ; 0 ) \delta( x_0 - \xds ) \delta( x_{P / 2} - \xds )
  \tilde{R} ( \lambda )
\EE
with
\BE
  \tilde{R} ( \lambda ) = v_0 ( \lambda ) v_{P / 2} ( - \lambda ) R ( \lambda )
  ,
\EE
where $W ( x_1, \ldots, x_P ; 0 )$ and $R ( \lambda )$ has the same definition
as in \Sec{corrfunc}. Thus, the time derivative of $\Cff( t )$ can be
obtained as
\BE
  \frac{ \Cffddot( 0 ) }{ \Cdd( 0 ) }
  \simeq
   -
  \frac{1}{\hbar^2}
  \left\langle
    \frac{ \tilde{R} ( \delta \lambda ) + \tilde{R} ( - \delta \lambda )
           - 2 \tilde{R}( 0 )
         }
         {
           ( \delta \lambda )^2
         }
  \right\rangle
  .
\EE
Similarly, virial estimators for higher time derivatives, $d^n
\Cff( 0 ) / d t^n ( n = 4, 6, \ldots )$, can be generated using
an appropriate finite-difference formula of higher order.\cite{PredescuRate}


\subsection{Global-path approach
\label{Sec:path_global}}

One can also devise an alternate ``global-path'' estimator
by first performing the coordinate scaling and then applying the flux
operator (i.e., in an opposite order to the preceding section).
To be specific, we insert the coordinate representation of the flux operator,
\BE
  \Fop = \frac{\hbar}{2 m i}
  \int dx
  \left[
    - |x \acute{\rangle} \langle x|
    + |x \rangle \grave{\langle} x|
  \right]
\EE
with $|x \acute{\rangle} = \partial |x \rangle / \partial x$ into the
imaginary-time flux correlation function as
\BEA
  \Cbarff ( \lambda )
  & = &
  \left( \frac{\hbar}{2 m} \right)^2
  \int d x_a \int d x_b \delta( x_a - \xds ) \delta( x_b - \xds )
  \nonumber
  \\
  &  & \times
  \left\{
    \grave{\langle} x_a | e^{- ( \beta / 2 - \lambda ) \Hop}
    |x_b \acute{\rangle} \langle x_b | e^{- ( \beta / 2 + \lambda ) \Hop}
    | x_a \rangle
  \right.
  \nonumber
  \\
  &  & + \langle x_a | e^{- ( \beta / 2 - \lambda ) \Hop} | x_b \rangle
  \grave{\langle} x_b | e^{- ( \beta / 2 + \lambda ) \Hop} | x_a \acute{\rangle}
  \nonumber
  \\
  &  & - \grave{\langle} x_a | e^{- ( \beta / 2 - \lambda ) \Hop} | x_b \rangle
  \grave{\langle} x_b | e^{- ( \beta / 2 + \lambda ) \Hop} |x_a \rangle
  \nonumber
  \\
  &  &
  \left.
    - \langle x_a | e^{- ( \beta / 2 - \lambda ) \Hop} |x_b
    \acute{\rangle} \langle x_b | e^{- ( \beta / 2 + \lambda ) \Hop} |x_a
    \acute{\rangle}
  \right\}
  ,
\EEA
which can be written more compactly as
\BEA
\label{Eq:Cbarff}
  \Cbarff ( \lambda )
  & = &
  \int d x_a \int d x_b \delta( x_a - \xds ) \delta( x_b - \xds )
  \nonumber
  \\
  &  &
  \times \mathcal{F}^2
  \langle
    x_a^- | e^{- ( \beta / 2 - \lambda ) \Hop}
    | x_b^- \rangle \langle x_b^+ | e^{- ( \beta / 2 + \lambda ) \Hop} | x_a^+
  \rangle
  ,
\EEA
where an operator representing the ``square`` of the flux operator is given by
\BE
  \mathcal{F}^2
  = \left( \frac{\hbar}{2 m} \right)^2
  \lim_{x_a^{\pm} \rightarrow x_a}
  \lim_{x_b^{\pm} \rightarrow x_b}
  \left\{
    \frac{\partial^2}{\partial x_a^+ \partial x_b^+} +
    \frac{\partial^2}{\partial x_a^- \partial x_b^-} -
    \frac{\partial^2}{\partial x_a^+ \partial x_b^-} -
    \frac{\partial^2}{\partial x_a^- \partial x_b^+}
  \right\}
  .
\EE
Next we use a generalized scaling relation of the form (see Appendix):
\BE
\label{Eq:gen_scaling}
  \langle
    x_b' | e^{- \beta' \Hop} | x_a'
  \rangle
  =
  \int d x_1 \cdots \int d x_{N - 1}
  W_{ba} ( x_1, \ldots, x_{N - 1} ; \beta )
  R_{ba}^{\#} ( \beta' )
  ,
\EE
where
\BE
  R_{ba}^{\#} ( \beta' )
  =
  \frac{ \langle x_b' | e^{- \beta' \Top} | x_a' \rangle}
       { \langle x_b  | e^{- \beta  \Top} | x_a  \rangle}
  \exp
  \left\{
    - \frac{1}{N} \sum_{k = 0}^N w_k
    \left[
      \beta' V ( x_k' ) - \beta V ( x_k )
    \right]
  \right\}
  ,
\EE
and
\BEA
  x_k'
  & = &
  \bar{x}_k' + \sqrt{\frac{\beta'}{\beta}} ( x_k - \xref_k )
  ,
  \\
  \bar{x}_k'
  & = &
  x_a' + ( x_b' - x_a' ) \frac{k}{N}
  .
\EEA
Other quantities such as $W_{ba}$ and $\xref_k$ are defined the same
as in \Sec{offdiag}. We note that the end-points $( x_a', x_b' )$ are included
in the coordinate transformation in addition to $\beta'$. Applying the above
relation to the density matrix elements in \Eq{Cbarff} with $N=P/2$ and
appropriate choice of end-points gives
\BE
  \Cbarff ( \lambda )
  =
  \int d x_1 \cdots \int d x_P \delta( x_0 - \xds ) \delta( x_{P / 2} - \xds )
  W ( x_1, \ldots, x_P ; 0 )
  \mathcal{F}^2
  R^{\#} ( x_a^{\pm}, x_b^{\pm}, \lambda )
  ,
\EE
where $R^{\#} ( x_a^{\pm}, x_b^{\pm}, \lambda ) = \Rkin^{\#} \Rpot^{\#}$ with
\BE
  \Rkin^{\#}
  =
  \frac{ \langle x_a^- | e^{- ( \beta / 2 - \lambda ) \Top} | x_b^- \rangle
         \langle x_b^+ | e^{- ( \beta / 2 + \lambda ) \Top} | x_a^+ \rangle
       }
       { \langle x_P | e^{- \beta \Top / 2} | x_{P / 2} \rangle
         \langle x_{P / 2} | e^{- \beta \Top / 2} | x_0 \rangle}
  ,
\EE
\BEA
  \Rpot^{\#}
  & = &
  \exp
  \left\{
    - \frac{1}{P} \sum_{k = 0}^{P / 2} \tilde{w}_k
    \left[
      ( \beta + 2 \lambda ) V ( x_k^{+'} ) - \beta V ( x_k )
    \right]
  \right.
  \nonumber
  \\
  &  &
  \left.
     - \frac{1}{P} \sum_{k = P / 2}^P \tilde{w}_k
     \left[
       ( \beta - 2 \lambda ) V ( x_k^{-'} ) - \beta V ( x_k )
     \right]
  \right\}
  ,
\EEA
and
\BSUB
\label{Eq:kth_coord}
\BEA
  x_k^{\pm'}
  & = &
  \bar{x}_k^{\pm} + \sqrt{ \frac{\beta \pm 2 \lambda}{\beta} }
  ( x_k - \xref_k )
  ,
  \\
  \bar{x}_k^+
  & = &
  x_a^+ + ( x_b^+ - x_a^+ ) \frac{k}{P / 2}
  ,
  \\
  \bar{x}_k^-
  & = &
  x_b^- + ( x_a^- - x_b^- ) \frac{k - P / 2}{P / 2}
  .
\EEA
\ESUB
The $k$th coordinate in \Eq{kth_coord} with plus and minus signs are defined for $0
\leq k \leq P / 2$ and $P / 2 \leq k \leq P$, respectively. The time
derivative of $\Cff ( t )$ can be obtained by differentiating
the factor $\mathcal{F}^2 R^{\#} ( x_a^{\pm}, x_b^{\pm},\lambda )$
with respect to $\lambda$, 
where the $\mathcal{F}^2$ operator is applied analytically
using up to second-order potential derivatives.\cite{PredescuRate}
The latter operation is costly but often
not too demanding because
$\mathcal{F}^2$ involves only the coordinates that define the (generalized)
reaction coordinate, e.g., only a few Cartesian coordinates
that describe the reacting atoms.


\section{Numerical tests
\label{Sec:test}}

We now apply the above estimators to a one-dimensional system
with the Eckart potential barrier,
\BE
  V ( x ) = V_0 \mathrm{sech}^2 ( x / a )
  ,
\EE
where $V_0=0.425$ eV, $a=0.734$ au, and the mass is 1060 au,
which corresponds roughly to the H+H$_2$ reaction. Table 1 lists the
statistical error of $\Cdd^{( 2 )} / \Cdd$ and
$\Cff^{( n )} / \Cdd$ ($n = 2, 4, 6$) obtained with 1
million path samples (note that the time arguments are always $t=0$ and
are omitted hereafter).
Three estimators are compared: the thermodynamic estimator, the local-path
virial estimator in \Sec{path_local},
and the global-path virial estimator in \Sec{path_global}.
The latter two differ only in the treatment of the flux operator.
The number of path variables used was $P = 8$ for $T =$1000 K and $P =$40
for $T =$200 K, which have a discretization error of $\sim$2 \% of the
exact ($P\rightarrow\infty$) value. The dividing surface was always set at the top of the barrier with
$\xds = 0$ in \Eq{flux_op}. The reader is referred to \Refs{QI_extension}{PredescuRate} on how these time derivatives can be used to improve the approximate rates.

We see from Table 1 that the virial estimators always exhibit a smaller
statistical error than the thermodynamic estimator, as expected. Between the two
virial estimators, the global-path version has a smaller variance than the local
one by using more information on the entire path. The exceedingly small errors
of the global-path estimator ($< 0.1$ \%) at $T =$ 1000 K are somewhat
fortuitous, because at this temperature the system is close to the
free-particle limit and the global-path estimator
becomes exact for a free particle irrespective of the number of path
variables.\cite{FreeParticleLimit}
This situation does not occur for the local-path virial estimator,
where the velocity factor in \Eq{velfac} must be averaged even for a free particle
to give the correct result.
Another important fact is that the variance of the
virial estimators is nearly independent of the order of time derivatives
in contrast to the thermodynamic estimator, which
agrees qualitatively with the previous study by Predescu for the same system
using a Fourier-like path integral.\cite{PredescuRate}

Figure 1 plots the statistical error of $\Cff^{( 2 )} / \Cdd$
and $\Cff^{( 6 )} / \Cdd$ at $T = $ 200 K as
a function of the number of path variables $P$.
The variance of the thermodynamic
estimator grows rapidly with $P$, and the growth rate is especially large
for $\Cff^{( 6 )}$.
The local-path virial estimator also exhibits an increasing variance,
which is caused by the appearance of $P$
in the numerator of the velocity factor in \Eq{velfac}.
The global-path virial estimator, on the other hand, has a nearly constant
variance regardless of the value of $P$, thus facilitating the systematic
convergence to the $P \rightarrow \infty$ limit.

Next we apply the present method to a model proton transfer reaction 
\cite{Borgis}
in a polar solvent, $\mathrm AH + B \rightarrow A^- + HB^+$,
where $A$, $H$, and $B$ represent a hydrogen-bonding complex
dissolved in liquid methyl chloride at $T=$ 250 K.
The details of the model is given in \Ref{Borgis}. Here we quantize
only the proton degree of freedom with $P=$ 40
and use the path integral Monte Carlo
(MC) scheme described in \Ref{QI_borgis}.
Figure 2 shows the convergence
of $\Cddddot/\Cdd$,  $\Cff^{(n)}/\Cdd$ ($n=$ 2,4)
as a function of MC cycles. In all cases the virial estimators
outperform the thermodynamic estimators in convergence rate.
In particular, the convergence
of $\Cddddot$ is very rapid when using the virial estimator,
which is beneficial in calculating the QI rate in \Eq{QIrateCdd}. 
On the other hand, the statistical error becomes larger for $\Cff^{(n)}/\Cdd$,
and it was difficult to converge
with 2 million path samples for $n \geq 6$. 
This is in contrast to the one-dimensional
Eckart barrier studied above, where the variance of the virial estimator
was nearly independent of the order of time derivatives.
Apart from differences in the dimensionality of the system,
the variance may be increased by stiff potential
walls in the solute potential (defined with
Morse-like functions),\cite{Borgis} because
the virial estimator for $\Cff^{(n)}$ depends implicitly on the
higher-order potential derivatives.
For example, the local-path and global-path virial estimators for $\Cff^{(6)}$
depend on 7th- and 8th-order potential derivatives, although
the numerical calculation by finite difference needs only the 
1st- and 2nd-order derivatives of the potential. 
It is not clear at present to what extent
this behavior is common for other potentials (including polynomial potentials).
Nevertheless, the fast convergence of 2nd time derivatives even for the present
stiff potential is very encouraging when considering future
applications of the QI theory to more complex chemical reactions in
condensed phases.


\section{Concluding remarks
\label{Sec:concl}}

Our main purpose in this paper has been to show
how a virial estimator for the time derivative of correlation
functions can be obtained straightforwardly via a coordinate
scaling procedure, and that the resulting estimator has an expected smaller
variance than the thermodynamic estimator. We have also presented
two methods for treating the flux operator, i.e., local-path
and global-path approaches, in which the latter has a smaller
variance. The second time derivative of $\Cdd(t)$ and $\Cff(t)$
are clearly the most important quantities for the QI rate
in \Eq{QIrateCdd} or in \Eq{QIrateCff}.
An open problem is how to best utilize the higher-order derivatives
in order to improve the accuracy of approximate rates.
While some progress has been made in this
direction,\cite{QI_extension,PredescuRate}
more studies would be useful if we consider the availability of
$\Cff^{(n)}(0)$ at least for systems with well-behaved potentials.


\begin{acknowledgments}
This work was supported by the Director, Office of Science,
Office of Basic Energy Sciences, Chemical Sciences, Geosciences,
and Biosciences Division, U.S. Department of Energy under Contract
No. DE-AC03-76SF00098 and by the National Science Foundation
Grant No. CHE-0345280.
We also acknowledge a generous allocation of 
supercomputing time from the National Energy Research Scientific Computing
Center (NERSC). T.Y. acknowledges
the Grant-in-Aid for Scientific Research from the Ministry of Education
and Science in Japan for support of this work, and also thanks Cristian
Predescu for stimulating discussions on path integral techniques.
\end{acknowledgments}


\appendix*

\section{Using the Feynman-Kac formula}

The scaled expression (\ref{Eq:rho_ba_scaled}) can also be obtained as follows.
Utilizing the integration variables
$\{y_k\}$ defined by
\BE
\label{Eq:scale1}
  x_k' = \xref_k + \sqrt{ \frac{\hbar^2 \beta'}{m} } y_k
  ,
\EE
one can transform \Eq{rho_ba_prime} as follows,
\BE
  \rho_{ba}( \beta' )
  =
  \langle x_b | e^{ -\beta' \Top } | x_a \rangle
  \mathbb{E} \exp
  \left\{
    - \frac{ \beta' }{ N }
    \sum_{k=0}^N w_k V( x_k' )
  \right\}
\EE
with
\BE
  \mathbb{E}(\cdots)
  =
  \frac{
         \int dy_1 \cdots \int dy_{N-1}
         \exp\left\{
               - \frac{N}{2} \sum_{k=1}^N ( y_k - y_{k-1} )^2
             \right\}
         (\cdots)
       }
       {
         \int dy_1 \cdots \int dy_{N-1}
         \exp\left\{
               - \frac{N}{2} \sum_{k=1}^N ( y_k - y_{k-1} )^2
             \right\}
       }
  ,  
\EE
which becomes the Feynman-Kac formula in the $N \rightarrow \infty$ limit
with $\{y_k\}$ representing the standard Brownian bridge. Rewriting the above
equation as
\BE
  \rho_{ba}( \beta' )
  =
  \langle x_b | e^{ -\beta \Top } | x_a \rangle
  \mathbb{E} \exp
  \left\{
    - \frac{ \beta }{ N }
    \sum_{k=0}^N w_k V( x_k )
  \right\}
  R_{ba}( \beta' )
  ,
\EE
where $R_{ba}( \beta' )$ is defined by \Eq{R_ba} and
\BE
\label{Eq:scale2}
  x_k = \xref_k + \sqrt{ \frac{\hbar^2 \beta}{m} } y_k
  ,
\EE
and changing integration variables from $\{ y_k \}$ to $\{ x_k \}$
results in \Eq{rho_ba_scaled}.
Combining \Eqs{scale1}{scale2}
gives the coordinate transformation in \Eq{coord_trans}.
Similar procedures can be used to obtain a generalized expression
in \Eq{gen_scaling}.


%
%

\newcommand{ \PR }[2]{Phys. Rev. #1 \textbf{#2}}
\newcommand{ \PRL  }[1]{Phys. Rev. Lett. \textbf{#1}}
\newcommand{ \RMP  }[1]{Rev. Mod. Phys. \textbf{#1}}

\newcommand{ \JCP  }[1]{J. Chem. Phys. \textbf{#1}}

\newcommand{ \JPC  }[1]{J. Phys. Chem. \textbf{#1}}
\newcommand{ \JPCA }[1]{J. Phys. Chem. A \textbf{#1}}
\newcommand{ \JPCB }[1]{J. Phys. Chem. B \textbf{#1}}
\newcommand{ \JACS }[1]{J. Am. Chem. Soc. \textbf{#1}}
\newcommand{ \ACR }[1]{Acc. Chem. Res. \textbf{#1}}

\newcommand{ \ARPC  }[1]{Annu. Rev. Phys. Chem. \textbf{#1}}

\newcommand{ \CPC  }[1]{Comput. Phys. Commun. \textbf{#1}}

\newcommand{ \CP   }[1]{Chem. Phys. \textbf{#1}}
\newcommand{ \CPL  }[1]{Chem. Phys. Lett. \textbf{#1}}

\newcommand{ \EJB }[1]{Eur. J. Biochem. \textbf{#1}}
\newcommand{ \Biochem }[1]{Biochemistry \textbf{#1}}
\newcommand{ \JBC }[1]{J. Biol. Chem. \textbf{#1}}
\newcommand{ \OpinionBio }[1]{Curr. Opinion Struct. Bio.(???) \textbf{#1}}

\newcommand{ \Nature }[1]{Nature (London) \textbf{#1}}
\newcommand{ \Science }[1]{Science \textbf{#1}}
\newcommand{ \PNAS }[1]{Proc. Natl. Acad. Sci. USA \textbf{#1}}

\newcommand{ \MP }[1]{Mol. Phys. \textbf{#1}}
\newcommand{ \JMolLiq }[1]{J. Mol. Liq. \textbf{#1}}

\newcommand{ \ibid }{\textit{ibid}. }



%



\clearpage

\begin{figure}

\caption{
Relative statistical error (\%) of the thermodynamic, local-path virial,
and global-path virial estimators for (a) $\Cff^{(2)}/\Cdd$ and
(b) $\Cff^{(6)}/\Cdd$ computed for the one-dimensional
Eckart barrier at $T=$ 200 K. $P$ is the number of path variables.
}

\caption{
Statistical convergence of the thermodynamic, local-path virial, and
global-path virial estimators for
(a) $\Cdd^{(2)}/\Cdd$,
(b) $\Cff^{(2)}/\Cdd$, and
(c) $\Cff^{(4)}/\Cdd$
computed for a model proton transfer reaction in a polar solvent.
}

\end{figure}


\clearpage

\begin{flushleft}Fig.~1 (a)\end{flushleft}
\begin{center}\includegraphics[width=15cm,clip]{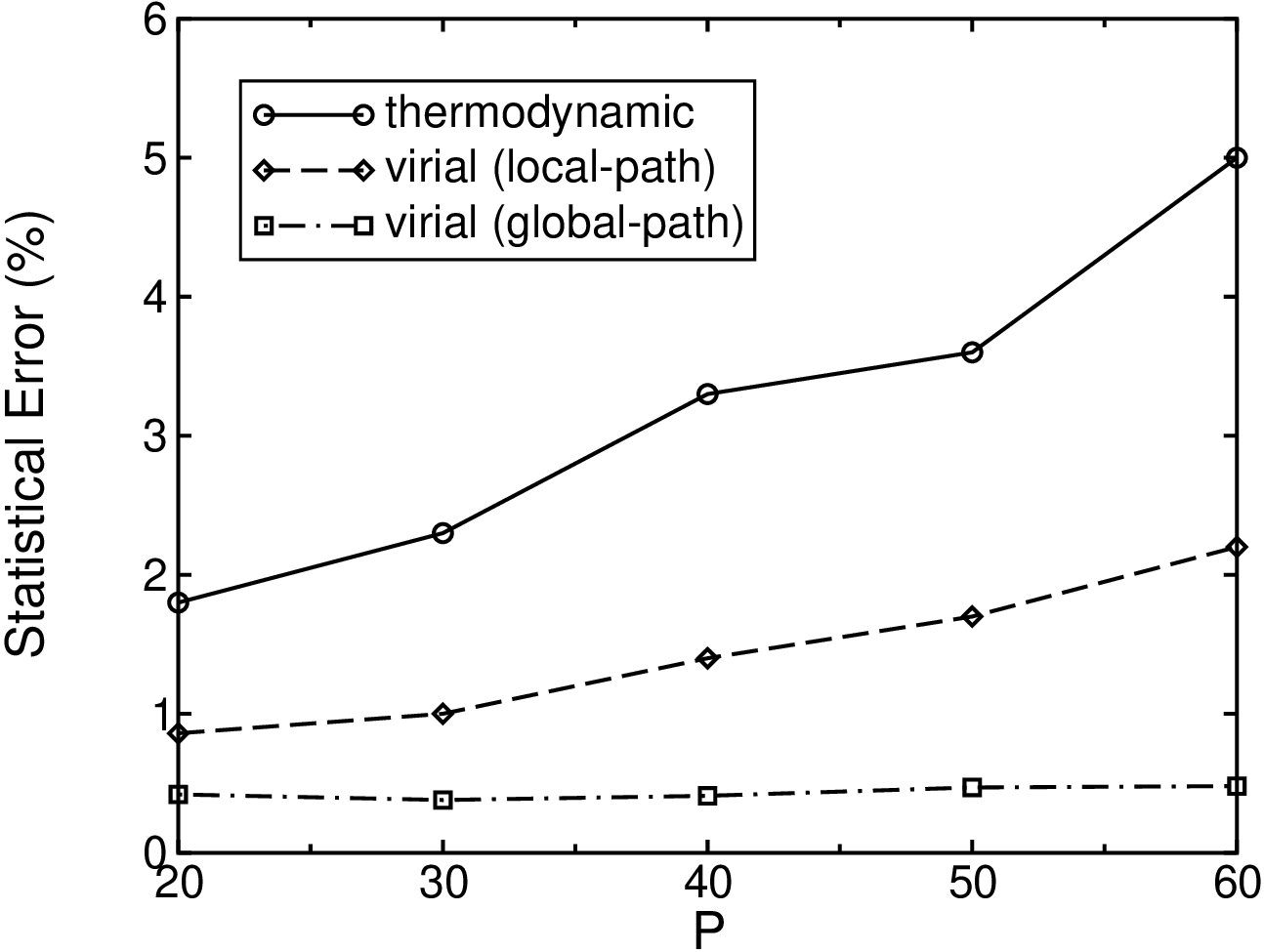}\end{center}

\clearpage

\begin{flushleft}Fig.~1 (b)\end{flushleft}
\begin{center}\includegraphics[width=15cm,clip]{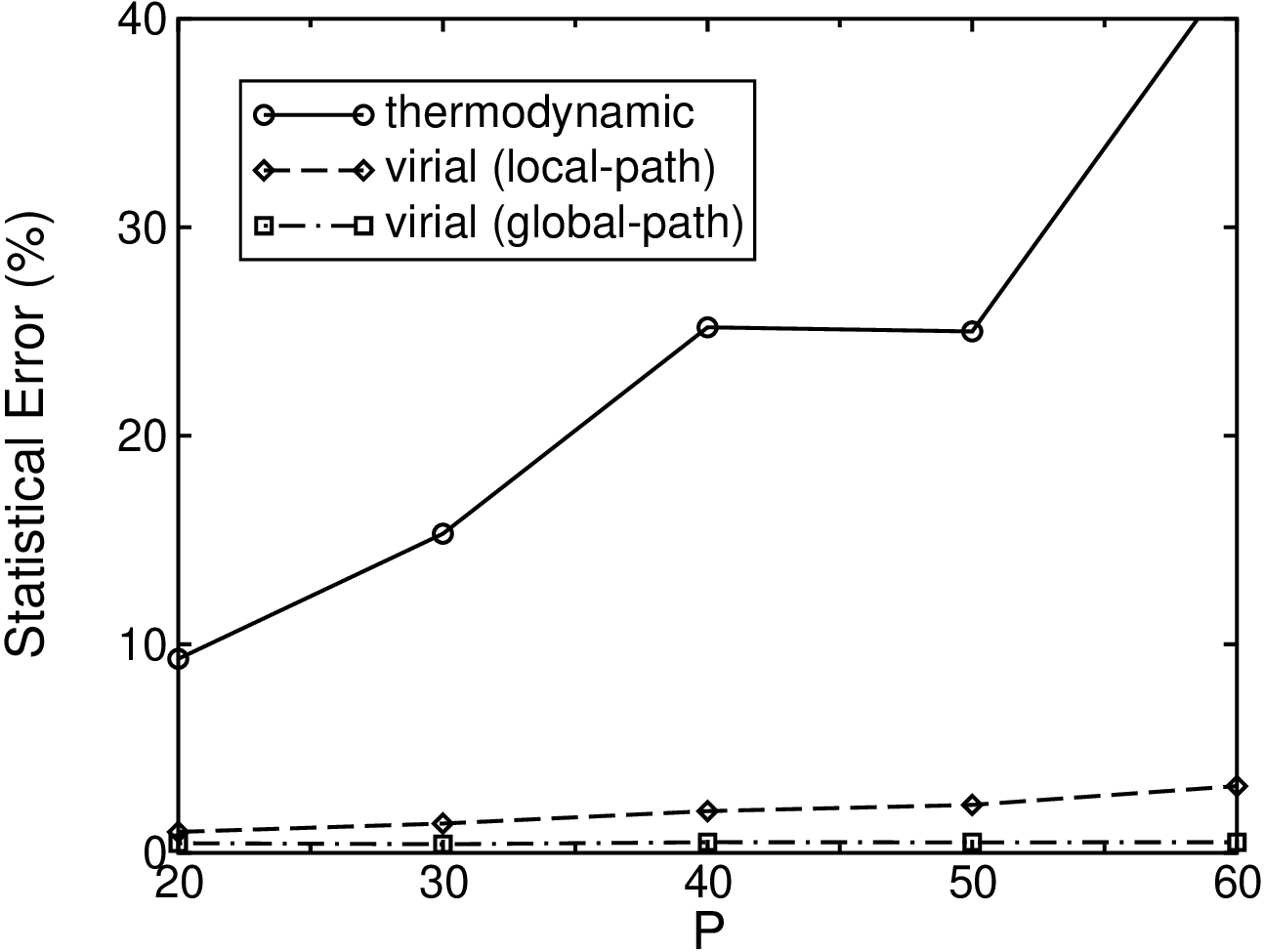}\end{center}


\clearpage

\begin{flushleft}Fig.~2 (a)\end{flushleft}
\begin{center}\includegraphics[width=15cm,clip]{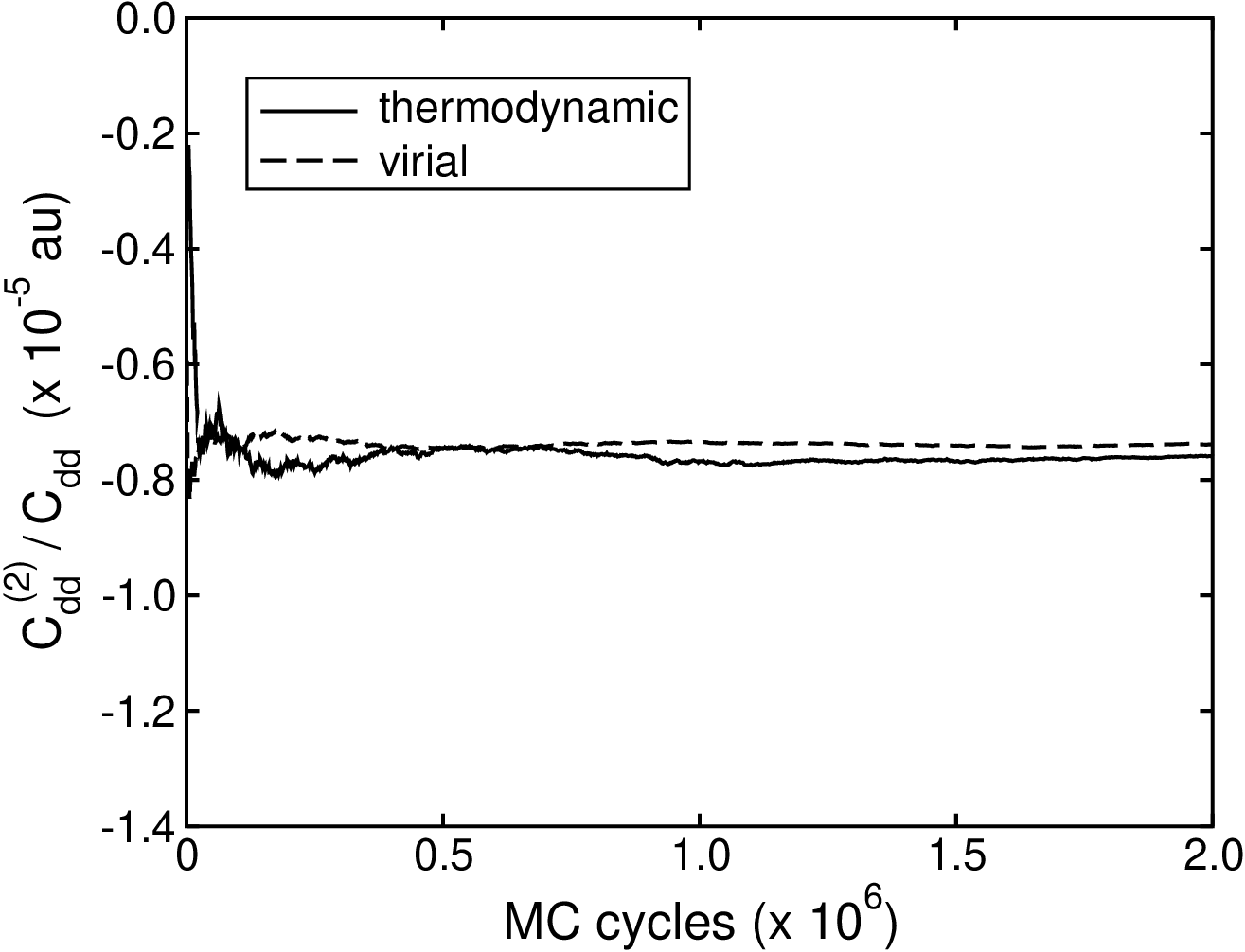}\end{center}

\clearpage

\begin{flushleft}Fig.~2 (b)\end{flushleft}
\begin{center}\includegraphics[width=15cm,clip]{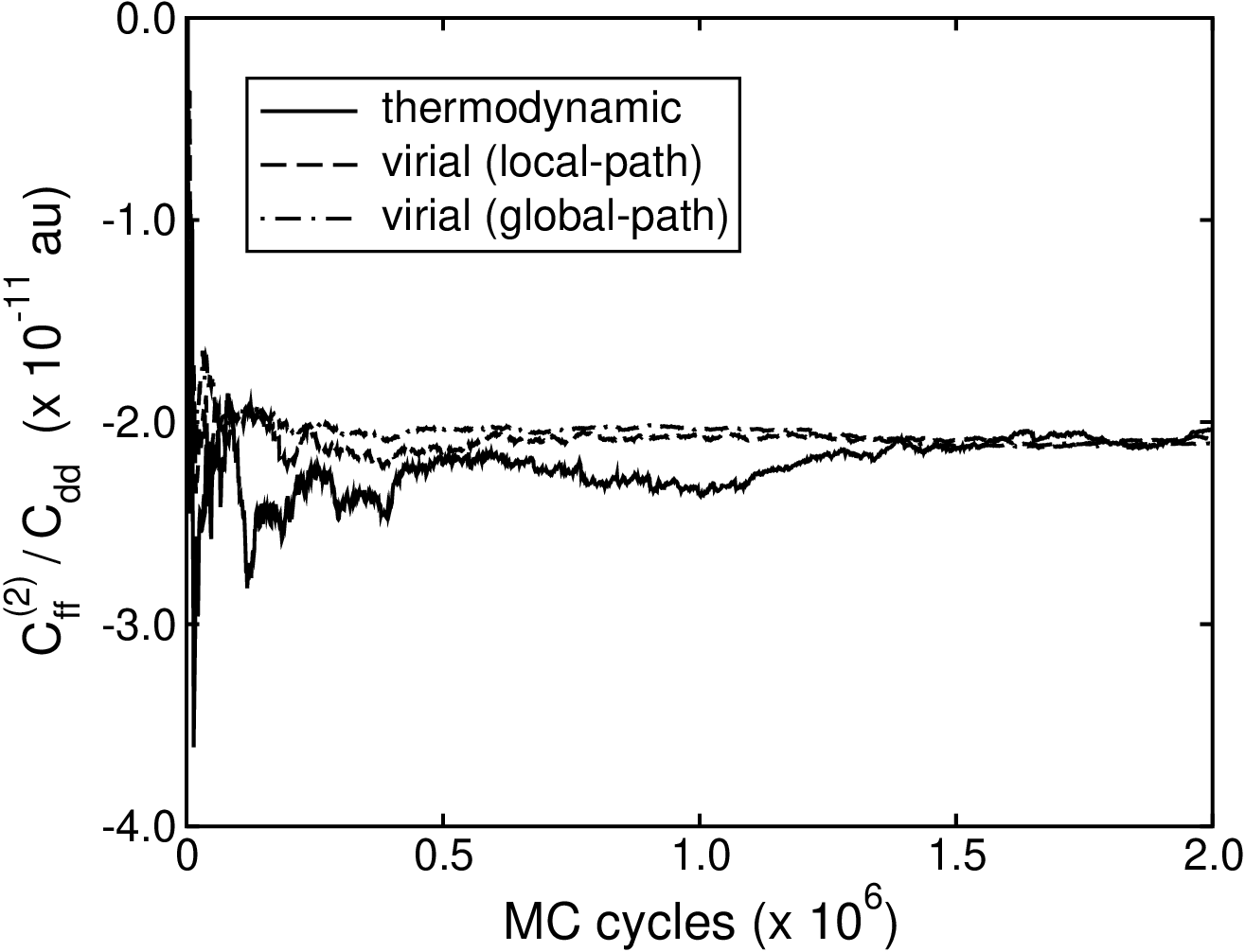}\end{center}

\clearpage

\begin{flushleft}Fig.~2 (c)\end{flushleft}
\begin{center}\includegraphics[width=15cm,clip]{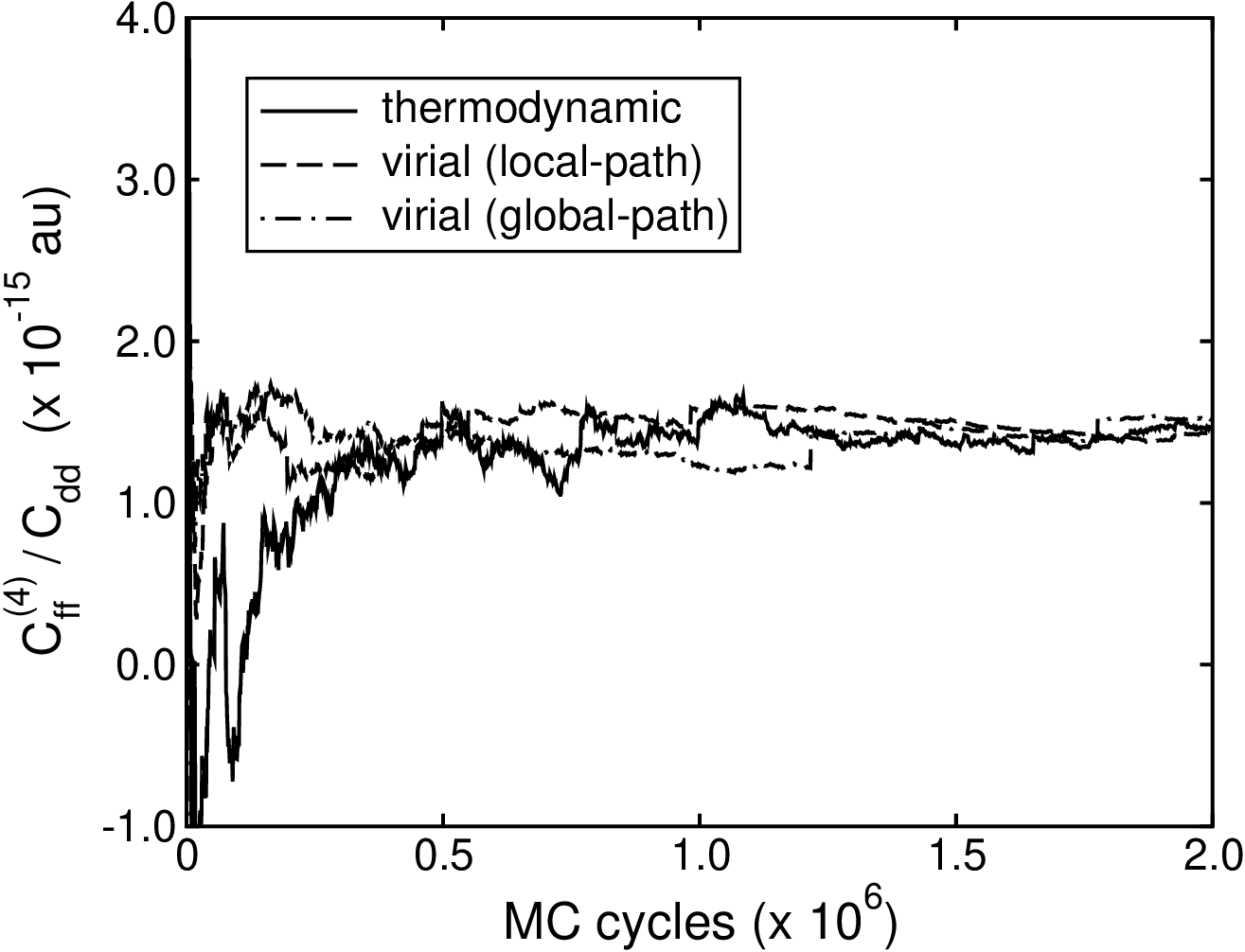}\end{center}



%


\clearpage

\begin{table}
\caption{\label{Table:eckart}
Relative statistical error (\%) of the thermodynamic, local-path virial,
and global-path virial estimators for the one-dimensional Eckart barrier.
One million paths are sampled with 8 and 40 path variables
for $T=$ 1000 and 200 K, respectively.
}
\begin{ruledtabular}
\begin{tabular}{cccc}
  &  thermodynamic  & virial (local-path)  & virial (global-path) \\
\hline
\multicolumn{4}{c}{$T=$ 1000 K} \\
$\Cdd^{(2)}/\Cdd$  &  0.5  &  0.024  &  0.024 \\
$\Cff^{(2)}/\Cdd$  &  1.2  &  0.34   &  0.012 \\
$\Cff^{(4)}/\Cdd$  &  2.3  &  0.33   &  0.014 \\
$\Cff^{(6)}/\Cdd$  &  2.6  &  0.33   &  0.015 \\
\hline
\multicolumn{4}{c}{$T=$ 200 K} \\
$\Cdd^{(2)}/\Cdd$  &  1.1  &  0.27  &  0.27 \\
$\Cff^{(2)}/\Cdd$  &  3.3  &  1.4   &  0.41 \\
$\Cff^{(4)}/\Cdd$  &  8.1  &  1.7   &  0.44 \\
$\Cff^{(6)}/\Cdd$  &  25   &  2.0   &  0.51 \\
\end{tabular}
\end{ruledtabular}
\end{table}




\bibliography{basename of .bib file}

\begin{thebibliography}{99}
\bibitem{FreeParticleLimit}
This is because if $V$ is constant, the correlation
function is determined solely from the free-particle contribution $\Rkin$.
\end{thebibliography}

\end{document}